# Comparative blobs and holes dynamics in a tokamak plasma: deep learning analysis of fast imaging data

**F Brochard[1*], H Aksoy[1], S Chouchène[1,2], J Cavalier[3], M Desecure[4] and N Lemoine[1]**

[1] Université de Lorraine, Institut Jean Lamour, CNRS, Nancy, France
[2] ICube, University of Strasbourg, France
[3] Institute of Plasma Physics of the CAS, Prague, Czech Republic
[4] APREX Solutions, Pulligny, France

[*]E-mail: frederic.brochard@univ-lorraine.fr

**Abstract.** This work focuses on the dynamics of the turbulent structures revealed by tomographic inversion of fast passive imaging data acquired on the COMPASS tokamak. To highlight the fluctuations, a sliding median image is subtracted from each image, revealing positive and negative structures. Assuming that the positive structures are blobs and the negative structures are holes, a recently developed deep learning analysis method is used to compare the dynamics of the two types of structures. While the results obtained for the positive structures seem to be in line with the dynamics expected for blobs, contradictory results are obtained for the negative structures, since their dynamics are very similar to those of blobs whereas they should be opposite. Our work suggests that the majority of negative structures resulting from data pre-processing are artefacts produced by the latter. However, a basic approach that only retains supernumerary negative structures shows that the behaviour of the latter is consistent with that expected for holes, opening new perspectives for their investigation.

## 1. Context of the work

Magnetically confined plasmas exhibit a wide range of instabilities driven by underlying gradients in density, temperature, electric potential, or magnetic field. Through nonlinear couplings, these instabilities can interact and develop into turbulent regimes. In nuclear fusion plasmas, such as those produced in tokamaks, turbulence gives rise to cross-field particle and energy transport that greatly exceeds diffusive levels. Fast imaging of the plasma edge reveals elongated filamentary structures aligned with magnetic field lines, which are a clear manifestation of this transport [1, 2]. These filaments, commonly referred to as blobs, correspond to localized regions of enhanced density and pressure. The pressure gradient within the blob creates polarization via curvature and ∇B drifts, leading to an electric dipole potential across the blob [3]. This generates an E×B drift that propels the blob radially outward, i.e. toward the scrape-off layer. In addition, the dipole potential sets up a poloidal drift: the blob tends to move in the poloidal direction corresponding to the ion diamagnetic direction. Similarly, regions of locally lower density and pressure are called holes. In a hole, the polarization is opposite to that of a blob: the potential dipole flips sign. This reversal in the E×B dipole flow causes holes to drift radially inward, i.e.

toward the core and to propagate in the opposite poloidal direction compared with blobs. For that reason, holes are expected to significantly contribute to the inward transport of impurities [4, 5].

Analyzing blob dynamics from fast imaging data presents several challenges. First, light intensity I is often used as a proxy to qualitatively assess plasma density, although the relationship between these two quantities also depends on the neutral density $n_0$ and electron density and temperature $(n_e, T_e)$, with $I = n_0 . f(n_e T_e)$ [6]. Second, when only passive imaging is employed, relying solely on the natural plasma emission, the signal-to-noise ratio is typically low, and tomographic inversion of camera data is required to localize the structures in 3D space [7]. To improve localization and enhance the signal-to-noise ratio, cameras are generally focused on a restricted region where localized gas puffing is applied; however, concerns remain regarding the perturbations introduced by this technique [6, 8]. Finally, a high acquisition rate is required to resolve the fast blob dynamics, resulting in a very large number of frames per video. Dedicated analysis methods, either automated or semi-automated, are therefore necessary to exploit these data. This last point, which accounts for the fact that tokamak video databases remain largely underexploited, may nevertheless become an asset for the development of supervised learning–based analysis methods, provided that sufficiently large labelled datasets can be generated [9, 10].

We have recently developed such a method to investigate blob dynamics and quantify their mutual interactions using high-resolution passive fast imaging data from the COMPASS tokamak at IPP Prague [11]. In this contribution, we examine the possibility of applying this method to study hole dynamics as well.

## 2. Methodology

### *2.1 Experimental setup*

The data used in this article were acquired on the COMPASS tokamak in the autumn of 2020, during one of the last measurement campaigns before its decommissioning, using a Photron SA-Z fast camera. This contribution focuses on the analysis of a single L-mode discharge, identified as shot #20846, with a toroidal magnetic field equal to 1.15 Tesla, 182 kA plasma current and flat-top density $4.10^{19}$ $m^{-3}$. The measurements were carried out in passive imaging mode, at an acquisition rate of 1,008,000 frames per second with an exposure time of 680 ns. At this frame rate, the active part of the sensor is composed of 128x40 pixels, covering a region of interest (ROI) of approximately 25 x 8 cm, but only a fraction of this ROI is exploitable, the light emission being mostly concentrated at the immediate vicinity of the last close flux surface (LCFS). The neutral population required for visible light emission relies on passive gas puff used for fuelling the plasma and from the recycling at the wall. The very high acquisition rate was only made possible thanks to the high sensitivity of the camera and the simplicity of the optical system, consisting of two lenses mounted in a telescope-like configuration, focused on a mirror tilted to have the lines of sights as tangential as possible to the magnetic field lines. The latter condition helps improving the quality of the tomographic reconstruction. The optical setup and the sensitivity of the observations to gas puffing are presented in [10].

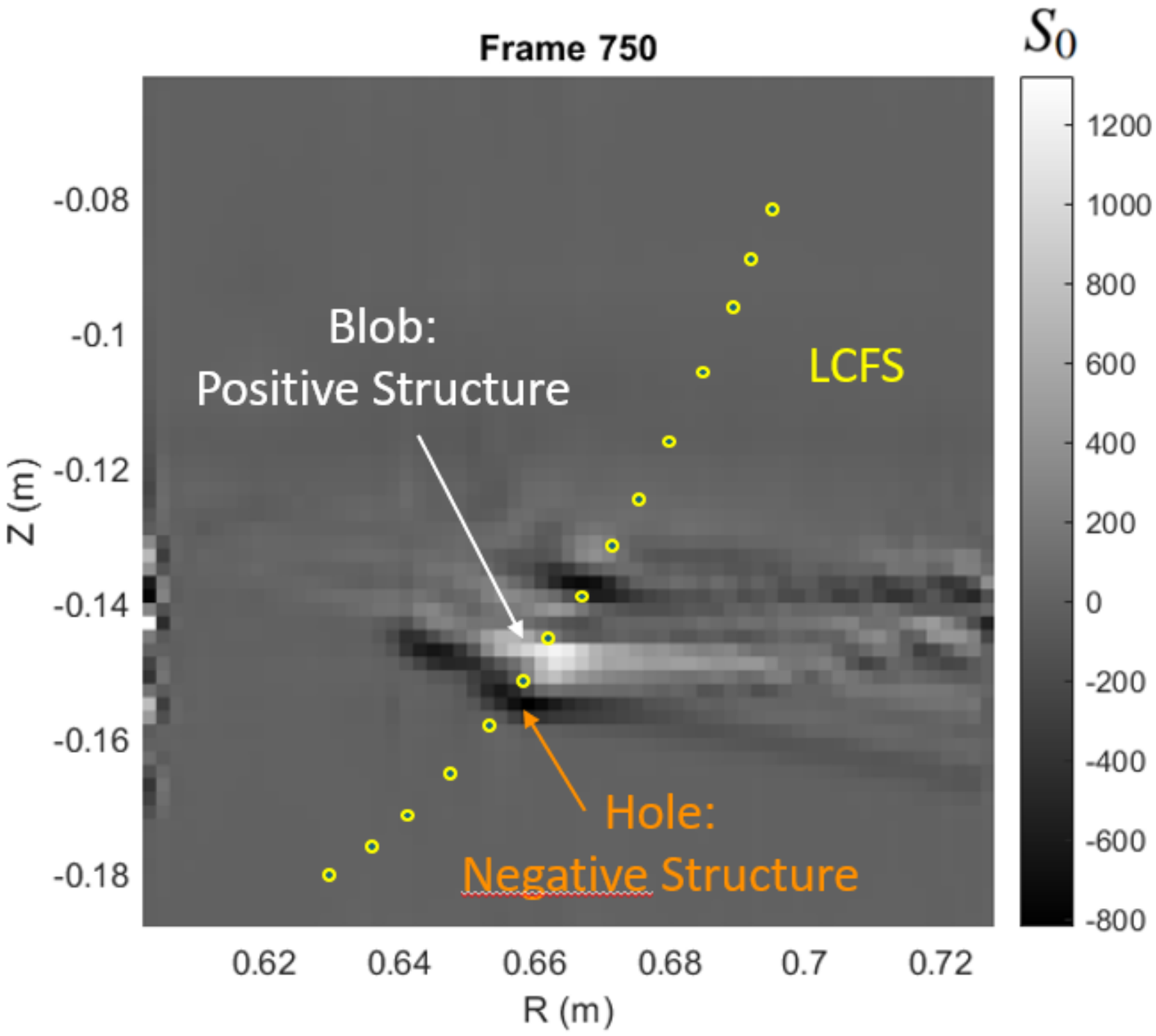

**Figure 1.** Emissivity map in the (R,Z) poloidal plane obtained after tomographic inversion and subtraction of a sliding temporal median image over 11 frames, revealing positive structures interpreted as blobs and negative structures interpreted as holes. The yellow dots depict the location of the Last Closed Flux Surface (LCFS). Some reconstruction artefacts are seen on the very left edge and on the right edge of the picture.

### *2.2 Data preparation*

In order to reconstruct the image of blobs in a poloidal cross-section, a tomographic inversion of the camera data is required. In other words, the inversion procedure aims to retrieve the local emissivity $S_0$ from the camera image $I_0$, which results from the integration of the plasma light emission along the camera's lines of sight:

$$I_0(x,y) = \int_{S_c}^{+\infty} S_0\left(\Psi(s_{xy}), \theta(s_{xy}), \varphi(s_{xy})\right) ds_{xy}$$

where (x,y) are the horizontal and vertical coordinates in the camera plane, $S_{xy}$ is the curvilinear abscissa along the ray passing through (x,y), and $S_c$ is the position of the pupil of the camera objective along this ray. The lines of sight are obtained by aligning the tokamak wall structures visible in low-frame-rate recordings with a CAD model of the tokamak interior, using the Calcam software [12]. The complete inversion procedure -based on the assumption of quasi-invariance of the emissivity along magnetic field lines in the volume imaged by the camera- is described in detail in [7].

The inversion requires precise knowledge of the position of the magnetic flux surfaces and verification of their stability over the considered time interval, in order to ensure accurate mapping of the camera data onto the magnetic geometry. For this purpose, we use the EFIT tool [13], which solves the Grad–Shafranov equation based on measurements from a set of diagnostics, with a verified average accuracy of 1 cm on the LCFS position [14]. If the magnetic surfaces reconstructed by EFIT are not sufficiently stable over the chosen time interval, the latter can be subdivided into shorter time windows, with a separate inversion performed for each. In the present case, a single inversion was performed for the time window (1100–1150 ms) of the discharge, corresponding to the discharge plateau where the EFIT reconstruction is stable. After inversion, 2D emissivity maps can be generated in any plane within the volume imaged by the camera, although the most accurate reconstruction is obtained when the focal plane is chosen. Figure 1 shows an example of the emissivity map in the poloidal plane closest to the focal plane, after subtraction of the median temporal background to highlight positive and negative structures, referred to as blobs and holes, respectively. To enhance the structures, while reducing the impact of the reconstruction artefacts at the edge of the map, an adaptive thresholding known as Bernsen filtering is ordinary applied before the next processing step.

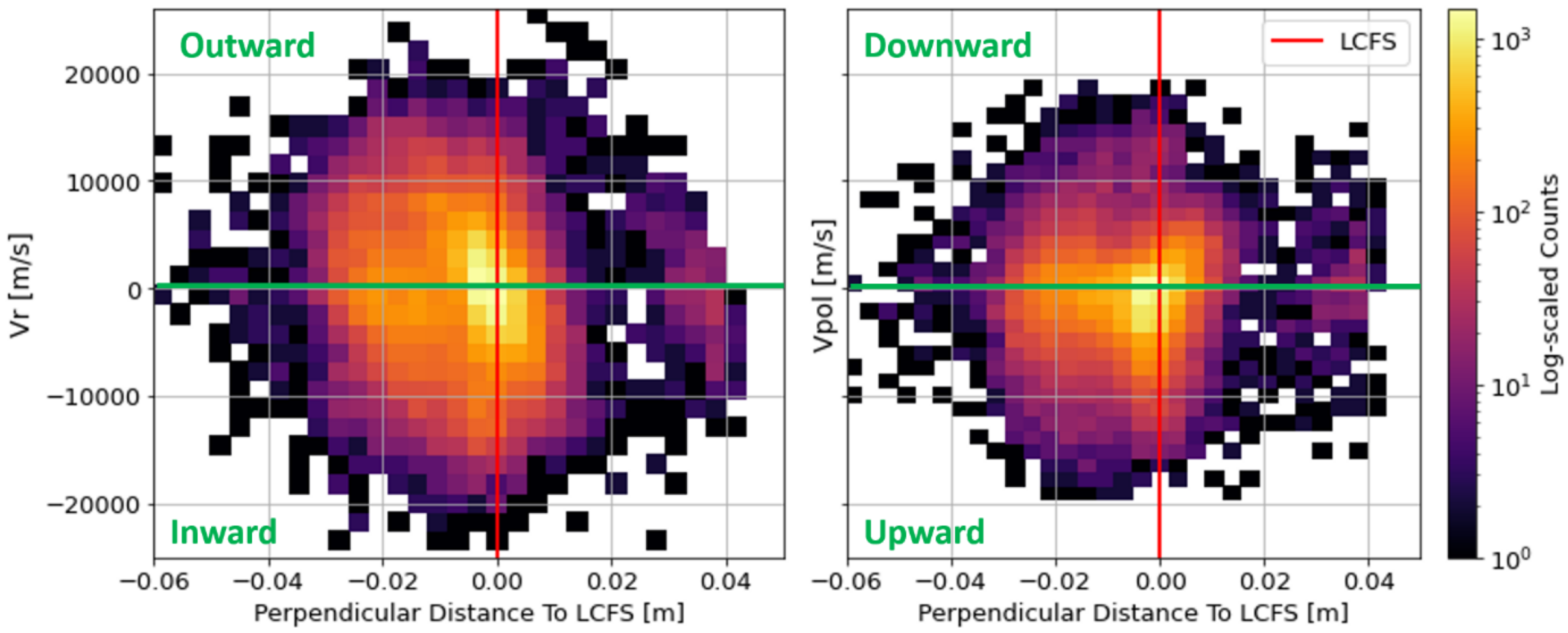

**Figure 2.** Heatmaps of the blob radial (left) and poloidal (right) velocities as a function of the distance to the LCFS (red vertical line), for discharge #20846.

Different types of analysis can then be performed on the reconstructed image sequences to study the dynamics of blobs and holes. In a previous work, we showed that deep learning improves the detection and tracking of blobs, and, for instance, allows us to accurately recover the position of the poloidal shear layer [10]. However, such a representation only imperfectly accounts for the complexity of the dynamics of the structures, whose velocity distribution varies strongly at each point. An illustration of this complexity is provided in Figure 2, which shows the distributions of the radial and poloidal velocities of blobs as a function of their distance from the LCFS. The counter-propagations of blobs resulting from turbulence in turn promote mutual interactions between blobs, which further sustain the turbulence. To characterize these interactions, we developed a method based on the study of the poloidal dynamics of turbulent structures, relying on the analysis of kymographs constructed according to the principle illustrated in Figure 3. The main advantage of the kymograph representation is that a single image enables the visualization of a large amount of information, which can then be easily extracted using pretrained convolutional neural networks. As illustrated in Figure 4, the analysis of patterns within the kymograph makes it possible to quickly identify five distinct types of behavior. Linear structures with a positive slope correspond to structures propagating poloidally upward (i.e. in the direction of increasing Z in Figure 3). Conversely, structures with a negative slope propagate downward. Multi-branched structures correspond to more complex dynamics. For example, a reversal of slope indicates a structure whose propagation direction changes. A branch splitting into two as time increases is the signature of a splitting event, while two branches merging together indicate the coalescence of two structures.

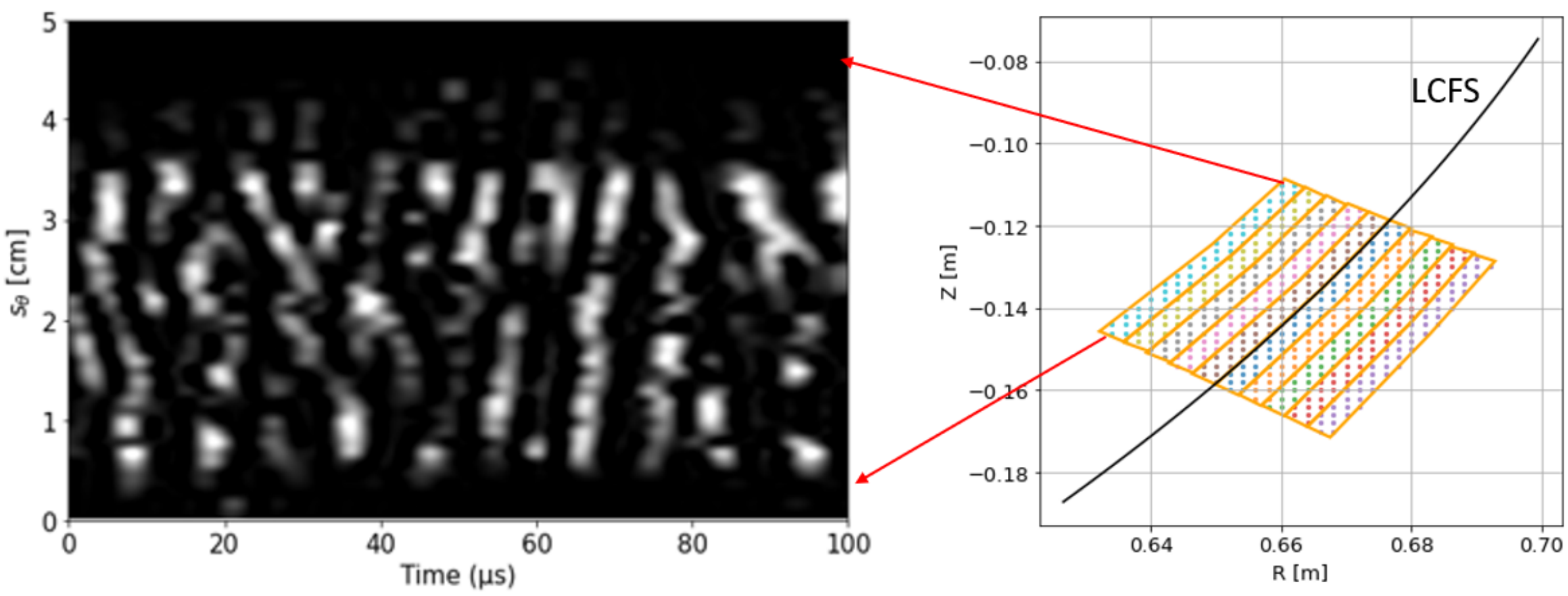

**Figure 3.** Principle of construction of poloidal kymographs. The 2D emissivity maps are divided into adjacent stripes oriented along constant flux surfaces of width 4mm (right). For each stripe, kymographs are constructed by depicting the locally average pixel intensity along the curvilinear abscissa as a function of time, resulting in a two-dimensional map of intensity as a function of poloidal position and time (left).

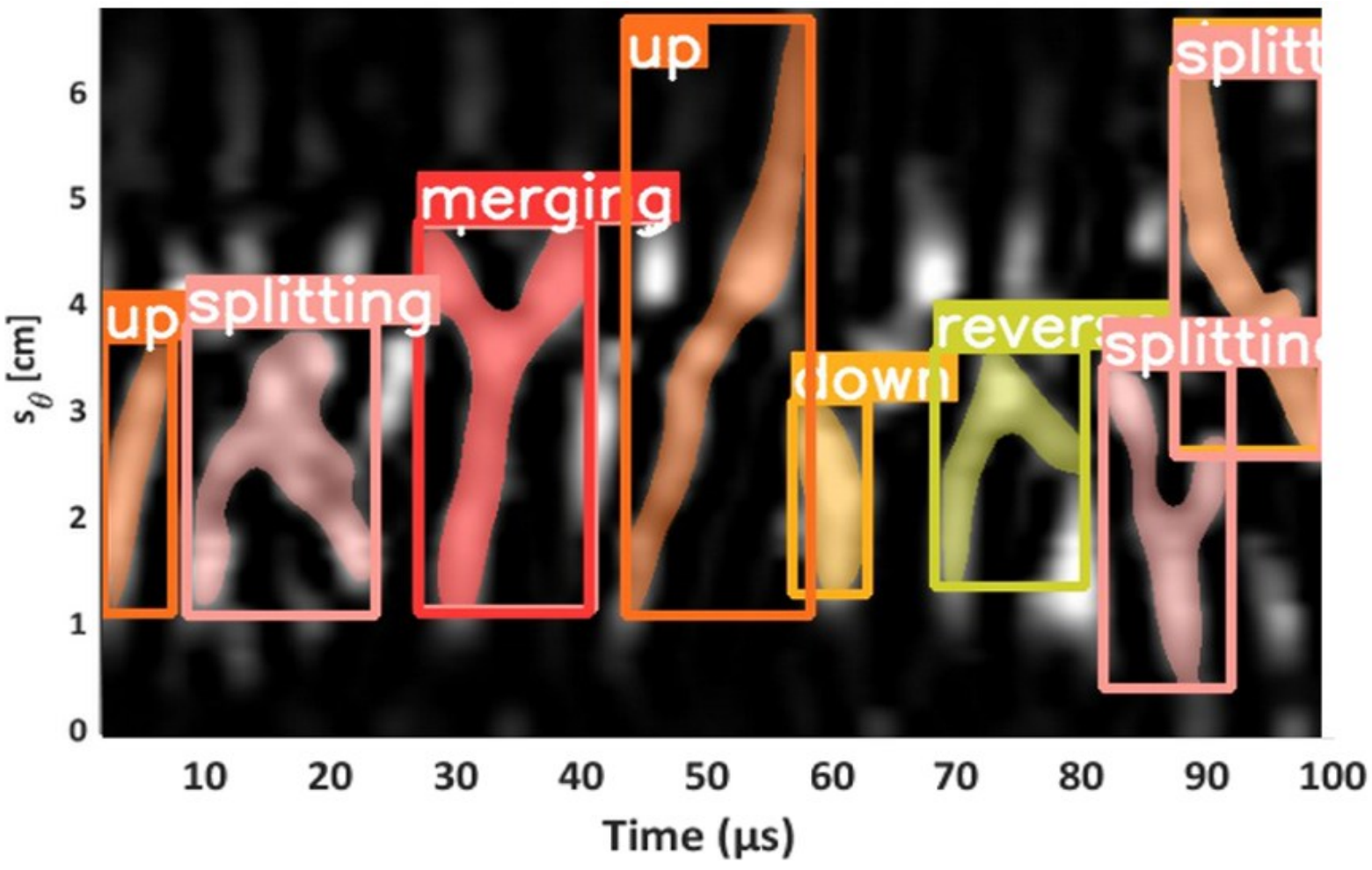

**Figure 4.** Labelled poloidal kymograph with the 5 classes of dynamical behaviour under investigation.

### *2.3 Deep Learning model*

Filament dynamics evidenced in kymographs are detected and classified using the deep learning model presented in [11], based on the YOLO (You Only Look Once) architecture [15], a widely used framework that processes an entire image in a single step to simultaneously predict the position and category of multiple objects. In our work, a segmentation variant, YOLO-Segmentation [16], is employed to simultaneously localize, classify, and delineate filamentary structures. The model was trained on 13,900 labeled samples from 2,300 manually annotated kymographs across three different discharges and adapted to identify five categories of blob motion: merging, splitting, upward, downward, and reversal. The motivation for adopting such a model lies in the limitations of conventional analysis methods, which often rely on subjective criteria and can vary with the operator or the selected parameters. In contrast, once the kymographs are prepared, the deep learning model enables a fully automated and objective detection process. While some bias may still arise during preprocessing—such as from filtering choices or the definition of flux surface size and position—the classification itself remains consistent and reproducible. This makes it possible to carry out large-scale analyses efficiently, without the variability that characterizes manual or semi-manual approaches.

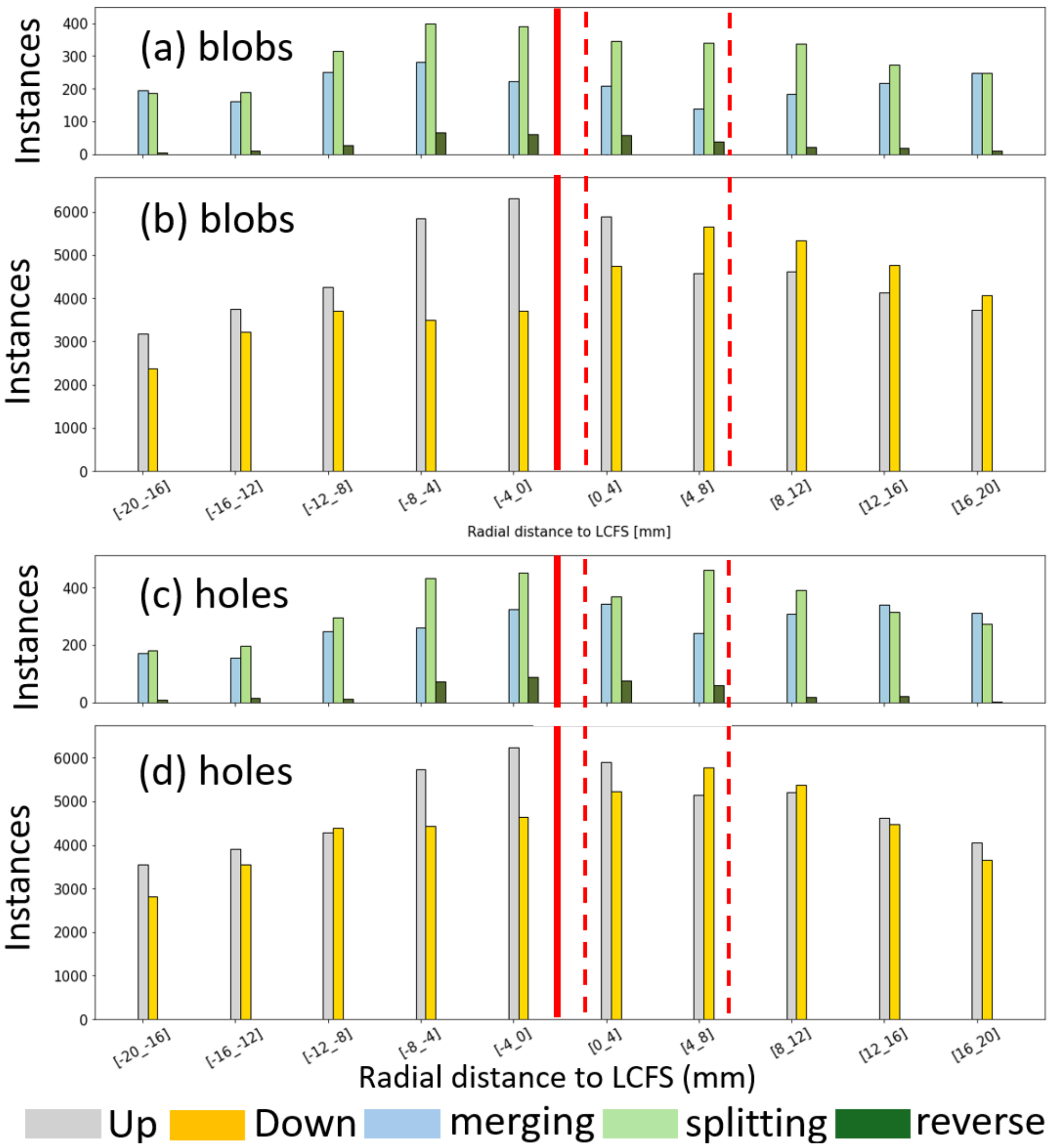

**Figure 5.** Number of detected events of the 5 classes of dynamical behaviour (upward, downward, merging, splitting, reversal) per 4mm radial bin as a function of the distance to the LCFS (red line). The dashed vertical lines indicate the location of the poloidal velocity shear layer obtained from 2D tracking.

## 3. Results and discussion

In the present article, the same model as the one used for investigating the blobs dynamics in [11] has been used, except that it is also used to investigate the dynamics of holes. For that purpose, the kymographs are built by taking the absolute value of the negative part of emissivity maps (cf. Figure 1). A synthetic view of the main results for both types of structures is given in Figure 5, which depicts the number of instances for each kind of dynamical behaviour as a function of the distance to the LCFS. It is striking that the results obtained for holes (Figure 5 c&d) are extremely close to those obtained for blobs (Figure 5 a&b). In both cases, the radial localization of the poloidal velocity shear layer agrees well with the inversion of the dominant number of instances corresponding to upward or downward displacements (Figure 5b&d). However, contrary to expectations, the overall motion proceeds in the same direction for both blobs and holes, except in the far SOL where a slight inversion is observed. For the other categories of dynamics (Figure 5a&c), the behavior is also very similar, although holes display a stronger tendency to coalesce within the shear layer and the SOL compared to blobs.

Overall, these results—too similar between two types of structures whose dynamics should in principle be opposite—questions the validity of our approach. In previous investigations, we showed that the reconstructed signal after tomographic inversion was relatively well correlated with ion saturation current measurements obtained from magnetically connected probes [7]. However, we cannot reproduce this investigation here, since the probes were not in operation during the recordings from the fast-imaging campaign dedicated to filament–filament interaction studies. The construction and analysis of kymographs require very high acquisition rates, typically above 700,000 fps, and we do not have datasets where both types of measurements are available simultaneously. At present, our main hypothesis is that, while positive emissivity peaks indeed correspond to blobs, negative values do not necessarily correspond to holes. We recall that the positive and negative values shown in Figure 1 are relative values, obtained after subtraction of a sliding median image to better reveal the structures. Subtracting the median value can therefore artificially create holes by a shadowing effect: when a blob moves, it leaves behind a void that may be mistakenly interpreted as a hole. As shown in Figure 5, the number of instances for holes is generally larger than for blobs. A simple approach to attempt retaining only the dynamics of genuine holes consists in subtracting the number of blob instances from that of the holes, as illustrated in Figure. 6. Although this approach is rather crude, the results obtained are very interesting and consistent with what is expected for genuine holes. In the confined region and up to the entrance of the LCFS, the dominant motion is directed downward, whereas for blobs it is directed upward. Beyond the LCFS and in the SOL, the dominant motion is reversed, oriented upward, while the dominant motion of blobs is downward. Furthermore, whereas dominant interactions typically lead to blob break-up, the dominant interaction for expected genuine holes more often results in merging. As in the previous figures, velocity reversals remain localized in the immediate vicinity of the shear layer.

In conclusion, this exploratory work suggests that fast imaging can be used to study not only blob dynamics but also hole dynamics in the vicinity of the LCFS. Particular care must be taken when tuning the median filter used to highlight the structures, since holes may be artificially created in large numbers by shadowing effects. Nevertheless, the analysis of such supernumerary holes yields interesting results that are consistent with the expected behaviour of genuine holes. To improve the method, we plan to investigate the influence of different preprocessing parameters, in particular the use of the median filter. This study will be based on comparisons between emissivity maps obtained with different parameter choices and measurements from magnetically connected probes. Such datasets are available in the COMPASS database, but at lower acquisition rates that are not suitable for analyses relying on the machine learning model developed for kymograph analysis. Instead, we plan to apply this model to study turbulent structure dynamics in the SPEKTRE device starting in 2026 [17].

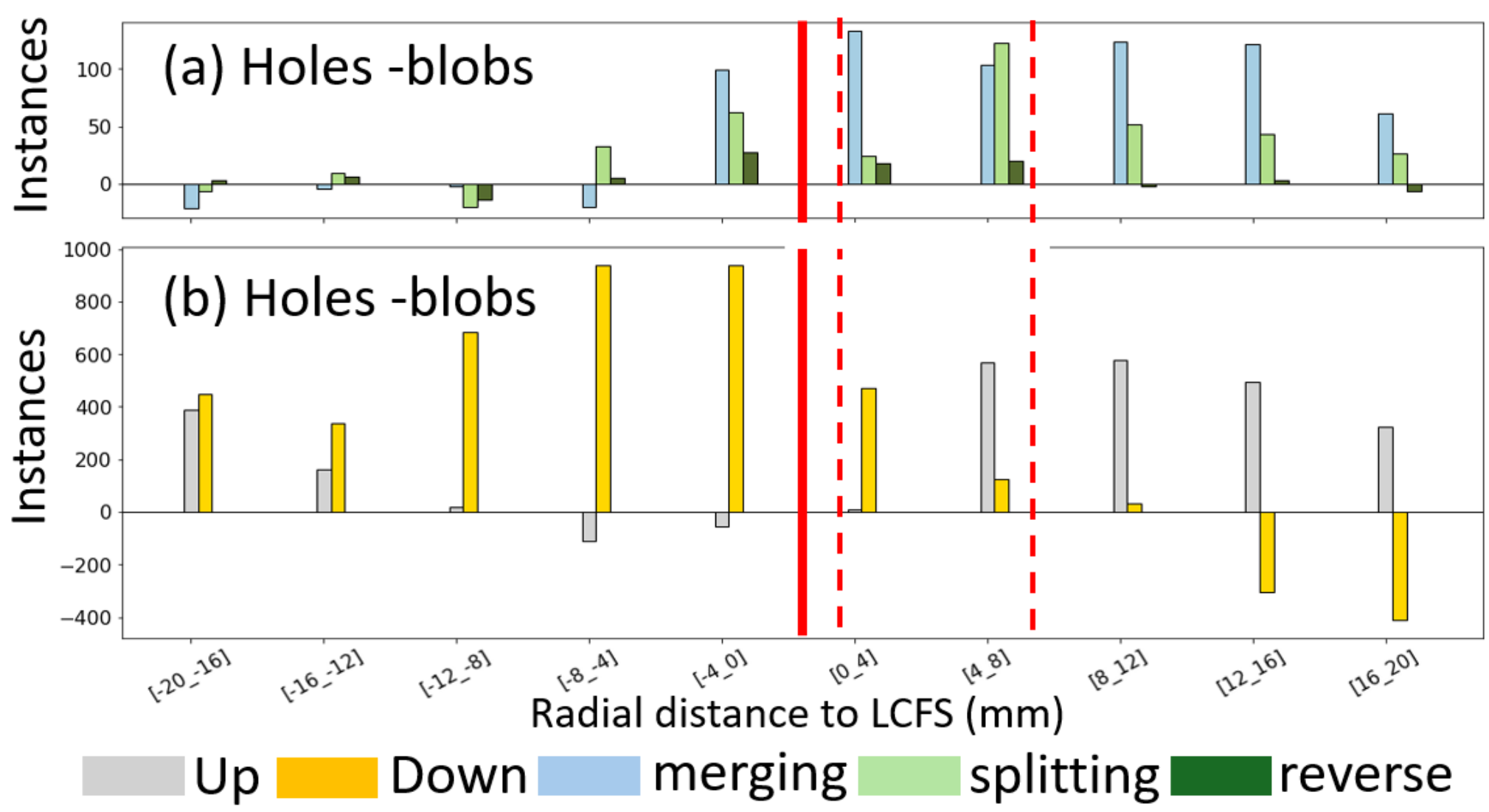

**Figure 6.** Difference in categorized dynamics between blobs and holes, computed by subtracting the hole counts from the corresponding blob counts for each radial bin. Positive values indicate a higher occurrence in hole dynamics, while negative values indicate a higher occurrence in blob dynamics. The location of the LCFS is indicated by the solid red line.

## Acknowledgements

This work has been carried out within the framework of the EUROfusion Consortium, funded by the European Union via the Euratom Research and Training Programme (Grant Agreement No 101052200 – EUROfusion). Views and opinions expressed are however those of the authors only and do not necessarily reflect those of the European Union or the European Commission. Neither the European Union nor the European Commission can be held responsible for them. This work has been co-funded by MEYS project # LM2023045. H. Aksoy would like to thank the Master Fusion-EP program (EMJMD), within which he contributed to this work during his internship.

## References

[1] Ben Ayed N, Kirk A, Dudson B, Tallents S, Vann R G L, Wilson H R and the MAST team 2009 *Plasma Phys. Control. Fusion* **51** 035016–40
[2] D'Ippolito D A, Myra J R and Zweben S J 2011 *Phys. Plasma* **18** 060501–49
[3] Krasheninnikov S I, D'Ippolito D A and Myra J R 2008 *J. Plasma Physics* **74** 679–717
[4] Krasheninnikov S I et al 2003 *Proceedings of the 19th IAEA Fusion Energy Conference*, Lyon, France, IAEA-CN-94/TH/4-1
[5] Hasegawa H and Ishiguro S 2017 *Nuclear Fusion* **57** 116008–15
[6] Zweben S J, Terry J L, Stotler D P and Maqueda R J 2017 *Rev. Sci. Instrum.* **88** 041101–22
[7] Cavalier J, Lemoine N, Brochard F, Weinzettl V, Seidl J, Silburn S, Tamain P, Dejarnac R, Adamek J and Panek R 2019 *Nucl. Fusion* **59** 056025–40
[8] Tamain P *et al* 2016 *Contrib. Plasma Phys.* **56** 569–74
[9] Han W, Pietersen R A, Villamor-Lora R, Beveridge M, Offeddu N, Golfinopoulos T, Theiler C, Terry J L, Marmar E S and Drori I 2022 *Scientific Reports* **12** 18142–51
[10] Chouchène S, Brochard F, Desecures M, Lemoine N, and Cavalier J 2024 *Scientific Reports* **14** 27965–80
[11] Chouchène S, Brochard F, Lemoine N, Cavalier J, Weinzettl V and Desecures M 2024 *Phys. Rev.* E. **109** 045201–10
[12] Silburn S *et al* Calcam DOI 10.5281/zenodo.1478554
[13] Brix M, Hawkes N C, Boboc A, Drozdov V, Sharapov S E and JET-EFDA Contributors 2008 *Rev. Sci. Instrum.* **79** 10F325

[14] Jirakova K, Kovanda O, Adamek J, Komm M and Seidl J 2019 *Journal of Instrumentation* **14** C11020
[15] Redmon J, Divvala S, Girshick R and Farhadi A 2015 arXiv:1506.02640
[16] Wang C Y, Bochkovskiy A and Liao H Y M 2022 arXiv:2207.02696
[17] Brochard F *et al* 2023 *Proceedings of the 49th EPS Conference on Plasma Physics*, Bordeaux, France